\newtheorem{theorem}{Theorem}
\newtheorem{lemma}{Lemma}
\newenvironment{proof}{{\sl Proof\/}:\ \ }{\qed\vspace{\baselineskip}}

\newcommand{\no}{\nonumber}
\newcommand{\Prob}{\textrm{Pr}}

\documentclass[conference]{IEEEtran}
\usepackage{times}
\usepackage[final]{graphicx}
\usepackage[reqno]{amsmath}
\usepackage{amsfonts}
\usepackage{times,amsmath,epsfig}
\usepackage{latexsym,amssymb}
\usepackage{cite}

\ifCLASSINFOpdf
\else
\fi
\begin{document}
%
\title{ARQ-Based Secret Key Sharing}

\author{\IEEEauthorblockN{Mohamed Abdel Ghany}
\IEEEauthorblockA{Wireless Intelligent Networks\\Center (WINC)\\Nile
University, Cairo, Egypt\\mohamed.abdelghany@nileu.edu.eg} \and
\IEEEauthorblockN{Ahmed Sultan} \IEEEauthorblockA{Wireless
Intelligent Networks\\Center (WINC)\\Nile University, Cairo,
Egypt\\asultan@nileuniversity.edu.eg} \and \IEEEauthorblockN{Hesham
El Gamal} \IEEEauthorblockA{Department of Electrical \\and Computer
Engineering\\Ohio State University, Columbus,
USA\\helgamal@ece.osu.edu}}


%


\maketitle

\begin{abstract}
This paper develops a novel framework for sharing secret keys using
existing Automatic Repeat reQuest (ARQ) protocols. Our approach
exploits the multi-path nature of the wireless environment to hide
the key from passive eavesdroppers. The proposed framework {\bf does
not assume} the availability of any prior channel state information
(CSI) and exploits only the one bit ACK/NACK feedback from the
legitimate receiver. Compared with earlier approaches, the main
innovation lies in the distribution of key bits among {\bf multiple}
ARQ frames. Interestingly, this idea allows for achieving a positive
secrecy rate even when the eavesdropper experiences more favorable
channel conditions, on average, than the legitimate receiver. In the
sequel, we characterize the information theoretic limits of the
proposed schemes, develop low complexity explicit implementations,
and conclude with numerical results that validate our theoretical
claims.

\end{abstract}

\section{Introduction}
Wireless communication, because of its broadcast nature, is
vulnerable to eavesdropping and other security attacks. Therefore,
pushing wireless networking to its full potential requires finding
solutions to its intrinsic security problems. In this paper, we
consider a physical layer-based scheme to share a secret key between
two users (Alice and Bob) communicating over a fading channel in the
presence of a {\bf passive eavesdropper} (Eve). The private key can
then be used to secure further exchange of information.

Arguably, the recent flurry of interest on wireless physical layer
secrecy was inspired by Wyner's wiretap
channel~\cite{Wyner1,Wyner2}. Under the assumption that Eve's
channel is a degraded version of Bob's, Wyner showed that
perfectly secure communication is possible by hiding the message
in the additional noise level seen by Eve. The effect of slow
fading on the secrecy capacity was studied later. In particular,
by appropriately distributing the message across different fading
realizations, it was shown that the multi-user diversity gain can
be harnessed to enhance the secrecy capacity,
e.g.~\cite{fading,NOCSI}. Another frame of work~\cite{Poor}
proposed using the well-known Hybrid ARQ protocols to facilitate the
exchange of secure messages between Alice and Bob.

This paper extends this line of work in two ways. First, by
distributing the key bits over multiple ARQ frames, we establish
the achievability of a vanishing probability of secrecy outages
~\cite{Poor} at the expense of a larger delay. Interestingly,
using this approach, a non-zero perfectly secure key rate is
achievable even when Eve is experiencing a more favorable average
signal-to-noise ratio (SNR) than Bob (unlike the scheme proposed
in~\cite{Poor}). Second, we develop explicit constructions for
secrecy ARQ coding that enjoy low implementation complexity. The
proposed scheme utilizes the ARQ protocol to create an erasure
wiretap channel and then uses known ideas from coset coding to
construct optimal codes for this channel
~\cite{LDPC1,LDPC2,LDPC3}.

The rest of this paper is organized as follows. Our system model
is detailed in Section~\ref{sec:sysmodel}.
Section~\ref{sec:infanal} provides the information theoretic
analysis of our model. Explicit secrecy coding schemes are
developed in Section~\ref{sec:imp}. In Section~\ref{sec:numeric},
we present numerical results. Finally, Section~\ref{sec:con}
summarizes our conclusions.

\section{System Model}\label{sec:sysmodel}
\begin{figure}
\centering
\includegraphics[width=0.45\textwidth,height=0.2\textheight]{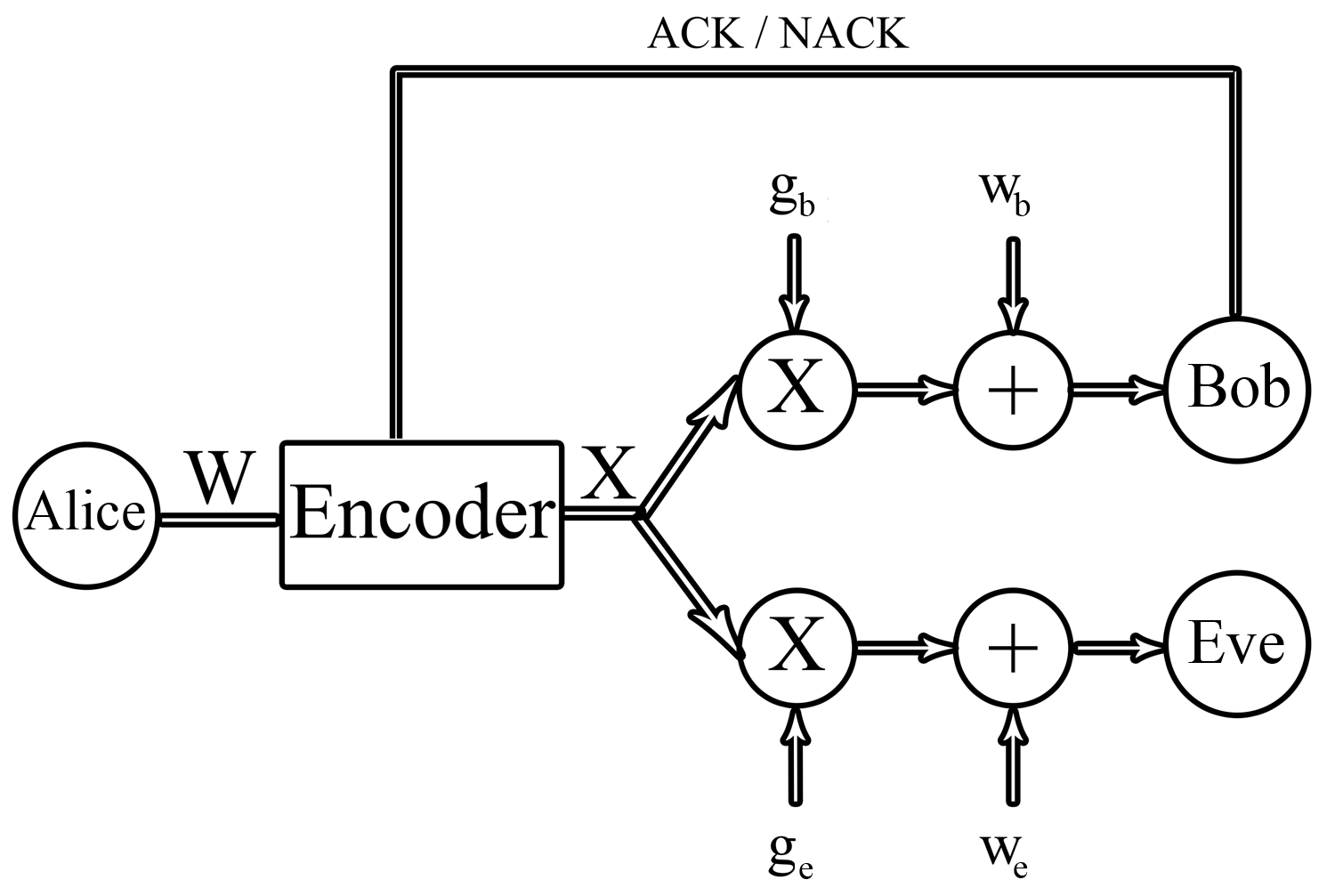}
\caption{System model involves a legitimate receiver, Bob, with a
feedback channel to the sender, Alice. Eve is a passive
eavesdropper. We assume block fading channels that are independent
of each other. \label{model}}
\end{figure}
Our model, shown in Figure~\ref{model}, assumes one transmitter (Alice), one legitimate receiver
(Bob) and one passive eavesdropper (Eve), all equipped with single
antenna. We adopt a block fading model in which the channel is
assumed to be fixed over one coherence interval and changes randomly
from one interval to the next. In order to obtain rigorous
information theoretic results, we consider the scenario of
asymptotically large coherence intervals and allow for sharing the
key across an asymptotically large number of those intervals. The
finite delay case will be considered in Section~\ref{sec:imp}. In
any particular interval, the signals received by Bob and Eve are
respectively given by,
\begin{eqnarray}
y(i,j)&=&g_b(j)\, x(i,j)+w_b(i,j),\\
z(i,j)&=&g_e(j)\, x(i,j)+w_e(i,j),
\end{eqnarray}
\noindent where $x(i,j)$ is the $\it {i}^{th}$ transmitted symbol
in the $\it {j}^{th}$ block, $y(i,j)$ is the $\it {i}^{th}$
received symbol by Bob in the $\it {j}^{th}$ block, $z(i,j)$ is
the $\it {i}^{th}$ received symbol by Eve in the $\it {j}^{th}$
block, $g_{b}(j)$ and $g_{e}(j)$ are the complex block channel
gains from Alice to Bob and Eve, respectively. Moreover,
$w_{b}(i,j)$ and $w_{e}(i,j)$ are the zero-mean, unit-variance
additive white complex Gaussian noise at Bob and Eve,
respectively. We denote the block fading power gains of the main
and eavesdropper channels by $h_b = |g_b(j)|^2$ and $h_e=
|g_e(j)|^2$. We do not assume any prior knowledge about the
channel state information at Alice. However, Bob is assumed to
know $g_b(j)$ and Eve is assumed to know both $g_b(j)$ and
$g_e(j)$. We impose the following short-term average power
constraint
\begin{eqnarray}
 {\mathbb E} \left(|x(i,j)|^2\right)\leq \bar{P}.
\end{eqnarray}

Our model only allows for one bit of ARQ feedback between Alice
and Bob. Each ARQ epoch is assumed to be contained in one
coherence interval (i.e., fixed channel gains) and that different
epochs correspond to independent coherence intervals (the same
assumptions as~\cite{Poor}). We denote the constant rate used in
each transmission frame by $R_0$ bits/channel use. The transmitted
packets are assumed to carry a perfect error detection mechanism
that Bob (and Eve) used to determine whether the packet has been
received correctly or not. Based on the error check, Bob sends
back to Alice an ACK/NACK bit, through a public and error-free
feedback channel. Eve is assumed to be passive (i.e., can not
transmit); an assumption which can be justified in several
practical settings. To minimize Bob's receiver complexity, we
adopt the memoryless decoding assumption implying that frames
received in error are discarded and not used to aid in future
decoding attempts.

\section{Information Theoretic Foundation}\label{sec:infanal}
In our setup, Alice wishes to share a secret key $W \in {\mathcal
W}=\{1,2,\cdots ,M\}$ with Bob. This key can be used for securing
future data transmission. To transmit this key, Alice and Bob use an
$(M,m)$ code consisting of : 1) a stochastic encoder $f_m(.)$ at Alice
that maps the key $w$ to a codeword $x^m \in {\mathcal X}^{m}$, 2) a
decoding function $\phi$: ${\mathcal Y}^{m}\rightarrow {\mathcal W}$
which is used by Bob to recover the key. The codeword is partitioned
into $a$ blocks each of $n_1$ symbols where $m = a\,n_1$. In this
section, we focus on the asymptotic scenario where
$a\rightarrow\infty$ and $n_1\rightarrow\infty$.

Alice starts with a random selection of the first block of $n_1$
symbols. Upon reception, Bob attempts to decode this block. If
successful, it sends an ACK bit to Alice who moves ahead and
makes a random choice of the second $n_1$ and sends it to Bob.
Here, Alice must make sure that the concatenation of the two
blocks belong to a valid codeword. As shown in the sequel, this
constraint is easily satisfied. If an error was detected, then Bob
sends a NACK bit to Alice. Here, we assume that the error
detection mechanism is perfect which is justified by the fact that
$n_1\rightarrow\infty$. In this case, Alice {\bf replaces} the
first block of $n_1$ symbols with another randomly chosen block
and transmits it. The process then repeats until Alice and Bob
agree on a sequence of $a$ blocks, each of length $n_1$ symbols,
corresponding to the key.

The code construction must allow for reliable decoding at Bob while
hiding the key from Eve. It is clear that the proposed protocol
exploits the error detection mechanism to make sure that both Alice and
Bob agree on the key (i.e., ensures reliable decoding). What remains
is the secrecy requirement which is measured by the equivocation
rate $R_e$ defined as the entropy rate of the transmitted key
conditioned on the intercepted ACKs or NACKs and the channel outputs
at Eve, i.e.,
\begin{equation}
R_e ~\overset{\Delta}{=}~ \frac{1}{n} H(W|Z^n,K^b,G_b^b,G_e^b) ~,
\end{equation} where $n$ is the number of symbols transmitted to exchange the key (including the symbols in the discarded blocks due to decoding
errors, $b=a\frac{n}{m}$, $K^b = \{ K(1), \cdots, K(b)\}$ denotes
sequence of ACK/NACK bits, $G_b^b$ and $G_e^b$ are the sequences of
channel coefficients seen by Bob and Eve in the $b$ blocks, and $Z^n
= \{ Z(1), \cdots, Z(n)\}$ denotes Eve's channel outputs in the $n$
symbol intervals. We limit our attention to the perfect secrecy
scenario, which requires the equivocation rate $R_e$ to be
arbitrarily close to the key rate. The secrecy rate $R_s$ is said to
be achievable if for any $\epsilon>0$, there exists a sequence of
codes $(2^{nR_s},m)$ such that for any $m\geq m(\epsilon)$, we have
\begin{equation}\label{secrecy}
R_e ~=~ \frac{1}{n} H(W|Z^n,K^b,G_b^b,G_e^b) ~\geq~ R_{s}-\epsilon
\end{equation}
and the key rate for a given input distribution is defined as the
maximum achievable perfect secrecy rate with this distribution. The
following result characterizes this rate, assuming a Gaussian input
distribution

\begin{theorem}\label{thm1}
For the memoryless ARQ , the perfect secrecy rate for {\bf
Gaussian inputs} for a given transmit power $P$ is given by:
\begin{equation}
C_s =  \max\limits_{R_0,P\leq \bar{P}}  \{\textrm{Pr}(R_0 \leq
\log(1+h_bP)) {\mathbb E}[R_0 - \log(1 + h_eP)]^{+}\},
\label{sec_cap}
\end{equation}
where $[x]^+=\max(0,x)$. All logarithms in this paper are taken to
base $2$, unless otherwise stated.
\end{theorem}
\begin{proof}
Here, we only give a sketch of the proof of achievability. Due to
space limitations, the converse will be deferred to the journal
paper version. The proof is given for a fixed average power $P\leq
\bar{P}$ and transmission rate $R_0$. The key rate is then obtained
by the appropriate maximization. Let $R_s = C_s - \delta$ for some
small $\delta
>0$ and $R = R_0 - \epsilon$. We first generate all binary sequences $\{
{\mathbf V} \}$ of length $m R$ and then independently assign each
of them randomly to one of $2^{n R_s}$ groups, according to a
uniform distribution. This ensures that any of the sequences are
equally likely to be within any of the groups. Each secret message
$w \in \{1, \cdots, 2^{n R_s} \}$ is then assigned a group ${\mathbf
V}(w)$. We then generate a Gaussian codebook consisting of $2^{n_1
\left(R_0 - \epsilon \right)}$ codewords, each of length $n_1$
symbols. The codebooks are then revealed to Alice, Bob, and Eve. To
transmit the codeword, Alice first selects a random group ${\mathbf
v}(i)$ of $n_1R$ bits, and then transmits the corresponding
codeword, drawn from the chosen Gaussian codebook. If Alice receives
an ACK bit from Bob, both are going to store this group of bits
and selects another group of bits to send in the next coherence
interval in the same manner. If a NACK was received, this group
of bits is discarded and another is generated in the same manner.
This process is repeated till both Alice and Bob have shared the
same key $w$ corresponding to $nR_s$ bits. We observe that the
channel coding theorem implies the existence of a Gaussian codebook
where the fraction of successfully decoded frames is given by
\begin{equation}
\frac{m}{n}=\textrm{Pr}(R_0 \leq \log(1+h_bP)),
\end{equation}
as $n_1\rightarrow\infty$. The equivocation rate at the eavesdropper can then be lower bounded as follows.
\begin{eqnarray}
n R_e &=& H(W|Z^n,K^n,G_b^b,G_e^b) \no \\
&\overset{(a)}{=}& H(W|Z^m,G_b^a,G_e^a) \no \\
&=& H(W,Z^m|G_b^a,G_e^a) - H(Z^m|G_b^a,G_e^a) \no \\
&=& H(W,Z^m,X^m|G_b^a,G_e^a) - H(Z^m|G_b^a,G_e^a) \no \\
&& \hspace{0.3in}- H(X^m|W,Z^m,G_b^a,G_e^a) \no \\
&=& H(X^m|G_b^a,G_e^a) + H(W,Z^m| X^m,G_b^a,G_e^a)  \no \\
&& \hspace{0.3in}- H(Z^m|G_b^a,G_e^a) - H(X^m|W,Z^m,G_b^a,G_e^a) \no \\
&\ge& H(X^m|G_b^a,G_e^a) + H(Z^m| X^m,G_b^a,G_e^a) \no \\
&& \hspace{0.3in}- H(Z^m|G_b^a,G_e^a) - H(X^m|W,Z^m,G_b^a,G_e^a) \no
\end{eqnarray}
\begin{eqnarray}
&=& H(X^m|G_b^a,G_e^a) - I(Z^m ; X^m|G_b^a,G_e^a)  \no \\
&& \hspace{0.3in}- H(X^m|W,Z^m,G_b^a,G_e^a)\no \\
&=& H(X^m | Z^m,G_b^a,G_e^a) - H(X^m|W,Z^m,G_b^a,G_e^a)\no\\
&\overset{(b)}{=}& \sum_{j=1}^a H(X(j)|Z(j),G_b(j),G_e(j)) \no \\
&& \hspace{0.3in}-H(X^m|W,Z^m,G_b^a,G_e^a)\no \\
&\overset{(c)}{\ge}& \sum_{j \in {\mathcal N}_m} H(X(j)|Z(j),G_b(j),G_e(j)) \no \\
&& \hspace{0.3in}-H(X^m|W,Z^m,G_b^a,G_e^a)\no \\
&=& \sum_{j \in {\mathcal N}_m} [ H(X(j)|G_b(j),G_e(j))  \no \\
&& \hspace{0.5in}- I(X(j);Z(j)|G_b(j),G_e(j)) ]  \no  \\
&& \hspace{1in}- H(X^m| W,Z^m,G_b^a,G_e^a) \no \\
&\ge & \sum_{j \in {\mathcal N}_m} n_1 \left[ R_0 - \log \left( 1 + h_e(j) P \right) - \epsilon \right]\no \no \\
&& \hspace{1in}- H(X^m|W,Z^m,G_b^a,G_e^a) \no \\
&\ge& \sum_{j=1}^{a} n_1 \left\{ \left[ R_0 - \log\left( 1 + h_e(j) P \right) \right]^{+} - \epsilon \right\} \no \\
&& \hspace{1in}- H(X^m|W,Z^m,G_b^a,G_e^a) \no \\
&\overset{(d)}{=}& n C_s - H(X^m|W,Z^m,G_b^a,G_e^a) - m \epsilon.
\label{lb1}
\end{eqnarray}
In the above derivation, (a) results from the independent choice of
the codeword symbols transmitted in each ARQ frame which does not
allow Eve to benefit from the observations corresponding to the
NACKed frames, (b) follows from the memoryless property of the
channel and the independence of the $X(j)$'s, (c) is obtained by
removing all those terms which correspond to the coherence intervals
$j \notin {\mathcal N}_m$, where ${\mathcal N}_m = \left\{ j \in
\{1, \cdots,a\} : h_b(j) > h_e(j) \right\}$, and (d) follows from
the ergodicity of the channel as $n, m \rightarrow \infty$. Now we
show that the term $H(X^m|W,Z^m,G_b^a,G_e^a)$ vanishes as ${n_1} \to
\infty$ by using a list decoding argument. In this list decoding, at
coherence interval $j$, the wiretapper first constructs a list
${\mathcal L}_j$ such that ${\bf x}(j) \in {\mathcal L}_j$ if $({\bf
x}(i),{\bf z}(i))$ are jointly typical. Let ${\mathcal L}={\mathcal
L}_1 \times{\mathcal L}_2\times\cdots\times{\mathcal L}_a$. Given
$w$, the wiretapper declares that $\hat{{\bf x}}^m=({\bf x}^m)$ was
transmitted, if $\hat{x}^m$ is the only codeword such that
$\hat{{\bf x}}^m \in B(w)\cap {\mathcal L}$, where $B(w)$ is the set
of codewords corresponding to the message $w$. If the wiretapper
finds none or more than one such sequence, then it declares an
error. Hence, there are two types of error events: 1) ${\mathcal
E}_1$: the transmitted codeword ${\bf x}^m_t$ is not in ${\mathcal
L}$, 2) ${\mathcal E}_2$: $\exists {\bf x}^m \neq {\bf x}^m_t$ such
that ${\bf x}^m \in B(w)\cap{\mathcal L}$. Thus the error
probability $\Prob (\hat{{\bf x}}^m \neq {\bf x}^m_t )= \Prob (
{\mathcal E}_1\cup {\mathcal E}_2 ) \leq \Prob ( {\mathcal E}_1) +
\Prob ({\mathcal E}_2)$. Based on the Asymptotic Equipartition
Property (AEP), we know that $\Prob ({\mathcal E}_1) \leq
\epsilon_1$. In order to bound $\Prob ({\mathcal E}_2)$, we first
bound the size of ${\mathcal L}_j$. We let
\begin{eqnarray}
\phi_j({\bf x}(j)|{\bf z}(j))=\left\{\begin{array}{ll}1,&
\textrm{$({\bf x}(j),{\bf z}(j))$ are jointly typical,} \\ 0,&
\textrm{otherwise.}\end{array} \right.
\end{eqnarray}
Now
\begin{eqnarray}
{\mathbb E}\{\|{\mathcal L}_j\|\}&=&{\mathbb
E}\left\{\sum\limits_{{\bf x}(j)}\phi_j(
{\bf x}(j)|{\bf z}(j))\right\}\no\\
&\leq&{\mathbb E}\left\{1+\sum\limits_{{\bf x}(j) \neq {\bf x}_t(j)}
\phi_j({\bf x}(j)|{\bf z}(j))\right\}\no\\
&\leq&1+\sum\limits_{{\bf x}(j) \neq {\bf x}_t(j)}{\mathbb
E}\left\{\phi_j({\bf x}(j)|
{\bf z}(j))\right\} \no\\
&\leq&1+2^{{n_1}\left[R_0 - \log(1+h_e(j)P)-\epsilon\right]}\no\\
&\leq&2^{{n_1}\left(\left[R_0-\log(1+h_e(j)P) - \epsilon \right]^+
+\frac{1}{n_1} \right)}
\end{eqnarray}
Hence
\begin{eqnarray}
{\mathbb E}\{\|{\mathcal L}\|\}=\prod\limits_{j=1}^{a} {\mathbb
e}\{\|{\mathcal L}_j\|\}=2^{\sum\limits_{j=1}^a n_1\left(\left[R_0-
\log(1+h_E(j)P) - \epsilon \right]^+ + \frac{1}{n_1}\right) }
\end{eqnarray}
\begin{eqnarray}
\Prob ({\mathcal E}_2 ) &\leq& {\mathbb E}\left\{\sum\limits_{x^m
\in{\mathcal L}, {\bf x}^m \neq {\bf x}^m_t}
\Prob ({\bf x}^m \in B(w)) \right\}\no\\
&\overset{(a)}\leq& {\mathbb E}\left\{\|{\mathcal
L}\|2^{-n R_s}\right\} \\
&\leq& 2^{-n R_s}2^{\sum\limits_{j=1}^a
n_1\left(\left[R_0-\log(1+h_e(j)P) - \epsilon \right]^+ +
\frac{1}{n_1} \right)
}\no\\
&\leq& 2^{-n \left(R_s -\frac{1}{c}\sum\limits_{j=1}^a
\left(\left[R_0
-\log(1+h_e(j)P) - \epsilon \right]^+ + \frac{1}{n_1} \right) \right)}, \no \\
&=& 2^{-n \left(R_s -\frac{1}{c}\sum\limits_{j=1}^a \left(\left[R_0
-\log(1+h_e(j)P) \right]^+ + \frac{1}{n_1} \right) +
\frac{|{\mathcal N}_m| \epsilon}{c} \right)}, \no
\end{eqnarray}
where (a) follows from the uniform distribution of the codewords in
$B(w)$. Now as $n_1 \to \infty$ and $a \to \infty$, we get \[ \Prob
({\mathcal E}_2 ) ~\le ~ 2^{-n \left(C_s - \delta - C_s + a \epsilon
\right)} ~=~ 2^{-n(c \epsilon -\delta)}, \] where $c = \Prob ( h_b >
h_e)$. Thus, by choosing $\epsilon > (\delta / c)$, the error
probability $\Prob ({\mathcal E}_2 ) \to 0$ as $n \to \infty$. Now
using Fano's inequality, we get \[ H(X^m|W,Z^m,G_b^a,G_e^a) ~\leq~ n
\delta_{n} \qquad \mbox{$\to 0$} \qquad \mbox{as  $m,n \to \infty$}.
\] Combining this with (\ref{lb1}), we get the desired result.
\end{proof}
A few remarks are now in order

\begin{enumerate}
\item It is intuitively pleasing that the secrecy key rate in
(\ref{sec_cap}) is the product of the probability of success at
Bob and the expected value of the additional mutual information
gleaned by Bob, as compared to Eve, in those successfully decoded
frames. \item It is clear from (\ref{sec_cap}) that a positive
secret key rate is achievable under very mild conditions on the
channels experienced by Bob and Eve. More precisely, unlike the
approach proposed in~\cite{Poor}, Theorem~\ref{thm1} establishes
the achievability of a positive perfect secrecy rate by
appropriately exploiting the ARQ feedback even when Eve's average
SNR is higher than that of Bob. \item Theorem~\ref{thm1}
characterizes the fundamental limit on secret key sharing and not
message transmission. The difference between the two scenarios
stems from the fact that the message is known to Alice {\bf
before} starting the transmission of the first block whereas Alice
and Bob can defer the agreement on the key till the last
successfully decoded block. This observation was exploited by our
approach in making Eve's observations of the frames discarded by
Bob, due to failure in decoding, useless. \item We stress the fact
that our approach does not require any prior knowledge about the
channel state information. The only assumption is that the public feedback channel is authenticated and only Bob can send over it.
\item The achievability of (\ref{sec_cap}) hinges on a random
binning argument which only establishes the existence of a coding
scheme that achieves the desired result. Our result, however,
stops short of explicitly finding such optimal coding scheme and
characterizing its encoding/decoding complexity. This observation
motivates the development of the explicit secrecy coding scheme in
the next section.

\item The perfect secrecy constraint imposed in (\ref{secrecy})
ensures that an eavesdropper with {\bf unlimited} computational
resources can not obtain any information about the key. In most
practical scenarios, however, the eavesdropper is only equipped
with limited computational power. The proposed scheme in the
following section leverages this fact in transforming our ARQ
secret sharing problem into an erasure-wiretap channel.
\end{enumerate}

\begin{figure}
\centering
\includegraphics[width=0.4\textwidth,height=0.095\textheight]{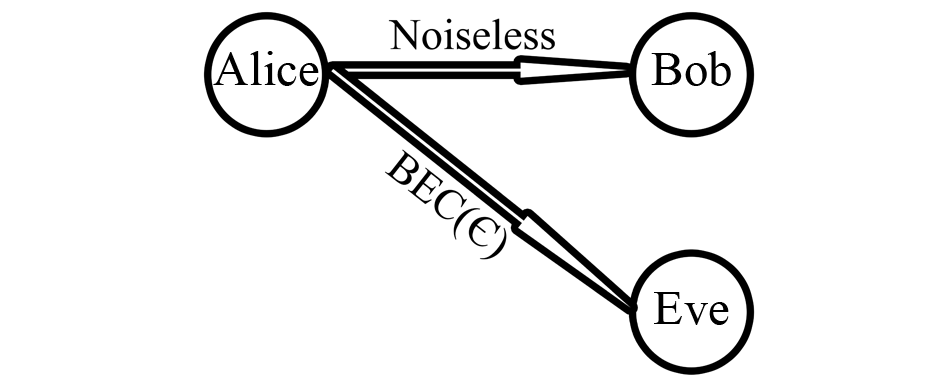}
\caption{Erasure-wiretap channel equivalent model. \label{BEC}}
\end{figure}

\section{Explicit Secrecy Coding Schemes}\label{sec:imp}

Inspired by the information theoretic results presented earlier,
this section develops explicit secrecy coding schemes that allow
for sharing keys using the underlying memoryless ARQ protocol. The
proposed schemes strive to minimize encoding/decoding complexity
at the expense of a minimal price in performance efficiency. We
proceed in three steps. The first step replaces the random binning
construction, used in the achievability proof of
Theorem~\ref{thm1}, with an explicit coset coding scheme for the
erasure-wiretap channel. As shown next, the erasure-wiretap
channel is created by the ACK/NACK feedback and accounts for the
computational complexity available to Eve. In the second step, we
limit the decoding delay by distributing the key bits over only a
finite number of ARQ frames. Finally, we replace the capacity
achieving Gaussian channel code with practical coding schemes in
the third step. Overall, our three-step approach allows for a nice
performance-vs-complexity tradeoff.

The perfect secrecy requirement used in the information theoretic
analysis does not impose any limits on Eve's decoding complexity. The
idea now is to exploit the finite complexity available at Eve in
simplifying the secrecy coding scheme. To illustrate the idea, let's
first assume that Eve can only afford maximum likelihood (ML)
decoding. Hence, successful decoding at Eve is only possible when

\begin{equation} R_0\leq \log(1+h_e P),\end{equation}
for a given transmit power level $P$. Now, using the idealized
error detection mechanism, Eve will be able to identify and erase
the frames decoded in error resulting in an erasure probability

\begin{equation} \epsilon=\textrm{Pr}(R_0 > \log(1+h_eP)).\label{ml}
\end{equation}

In practice, Eve may be able to go beyond the performance of the ML
decoder. For example, Eve can generate a list of candidate codewords
and then use the error detection mechanism, or other means, to
identify the correct one. In our setup, we quantify the
computational complexity of Eve by the amount of side information
$R_{\rm c}$ bits per channel use offered to it by a Genie. This side
information reduces the erasure probability to

\begin{equation} \epsilon_g=\textrm{Pr}(R_0-R_{\rm c}> \log(1+h_eP)),\label{rc}
\end{equation}
since now the channel has to supply only enough mutual information
to close the gap between the transmission rate $R_0$ and the side
information $R_{\rm c}$. The ML performance can be obtained as a
special case of (\ref{rc}) by setting $R_{\rm c}=0$.

It is now clear that using this idea we have transformed our ARQ
channel into an erasure-wiretap channel, as in Figure~\ref{BEC}. In this equivalent model,
we have a noiseless link between Alice and Bob, ensured by the
idealized error detection algorithm, and an erasure channel between
Alice and Eve. The following result characterizes the achievable
performance over this channel

\begin{lemma}\label{erasure1}
The secrecy capacity for the equivalent erasure-wiretap channel is
\begin{eqnarray}
C_e =  \max\limits_{R_0, P\leq \bar{P}}  \{R_0\textrm{Pr}(R_0 \leq
\log(1+h_bP)) \no\\ \textrm{Pr}(R_0-R_{\rm c} > \log(1+h_eP))\}.
\end{eqnarray}
\end{lemma}
The proof follows from the classical result on the erasure-wiretap
channel and is omitted here for brevity. It is intuitively
appealing that the expression in Lemma~\ref{erasure1} is simply
the product of the transmission rate per channel use, the
probability of successful decoding at Bob, and the probability of
erasure at Eve. The main advantage of this equivalent model is
that it lends itself to the explicit coset LDPC coding scheme
constructed in~\cite{LDPC1,LDPC2,LDPC3}. In summary, our first low
complexity construction is a concatenated coding scheme where the
outer code is a coset LDPC for secrecy and the inner one is a
capacity achieving Gaussian code. The underlying memoryless ARQ is
used to create the erasure-wiretap channel matched to this
concatenated coding scheme.

The second step is to limit the decoding delay resulting from the
distribution of key bits over an asymptotically large number of
ARQ blocks in the previous approach. To avoid this problem, we
limit the number of ARQ frames used by the key to a finite number
$k$. The implication for this choice is a non-vanishing value for
secrecy outage probability, which is the probability of Eve
obtaining correctly all $k$ frames. For example, if we encode the message as the syndrome of the rate $(k-1)/k$ parity check code then Eve will be completely blind about the key if {\em
at least} one of the $k$ ARQ frames is
erased ~\cite{LDPC1,LDPC2,LDPC3} (Here the distilled key is the modulo-$2$ sum of the key parts received correctly).
The secrecy outage probability is

\begin{equation}
P_{\rm out}= {\rm Pr}\left(\min\limits_{j\in\{1,...,k\}} \log
(1+h_{e}(j)P)> R_0 - R_{\rm c}\right),
\end{equation}
where $h_e(1)$,...,$h_e(k)$ are i.i.d. random variables drawn
according to the distribution of Eve's channel. Assuming a Rayleigh
fading distribution, we get

\begin{equation}
P_{\rm out}=\exp\left(-\frac{k}{P}\left[2^{R_0-R_{\rm
c}}-1\right]\right) \label{pout}.
\end{equation}

Under the same assumption, it is straightforward to
see that the average number of Bernoulli trials required to transfer
$k$ ARQ frames successfully to Bob is given by

\begin{equation}
N_0=k\exp\left(\frac{2^{R_0}-1}{P}\right), \label{avg_number}
\end{equation}

resulting in a key rate

\begin{equation}
R_k=\frac{R_0}{N_0}=\frac{R_0}{k}\exp\left(-\frac{2^{R_0}-1}{P}\right).
\end{equation}

Therefore, for a given $R_{\rm c}$ and $P$, one can obtain a
tradeoff between $P_{\rm out}$ and $R_k$ by varying $R_0$. Our
third, and final, step is to relax the assumption of a capacity
achieving inner code. Now, we allow for practical coding schemes,
including the possibility of uncoded transmission, with a finite
frame length $n_1$. Simulation results are reported in the next
section.

\section{Numerical Results}\label{sec:numeric}

Throughout this section we assume a Rayleigh fading channel, for
both Bob and Eve, and focus on the symmetric scenario where the
average SNRs experienced by both nodes are the same, i.e.,
${\mathbb E}\left(h_b\right)={\mathbb E}\left(h_e\right)$ = 1.
Under these assumptions, the achievable secrecy rate in
(\ref{sec_cap}) becomes
\begin{eqnarray}
\lefteqn{C_s= \max\limits_{R_0}
\exp\left(-\frac{2^{R_0}-1}{P}\right).{}}
\nonumber\\
& &
\left\{R_0-\frac{\exp\left(1/P\right)}{\log_e\left(2\right)}\left[E_{\rm
i}\left(1/P\right)-E_{\rm i}\left(2^{R_0}/P\right)\right]\right\}
\end{eqnarray}
\noindent where $E_{\rm
i}\left(x\right)=\int_{x}^{\infty}\exp\left(-t\right)/t\,dt$.

\begin{figure}
\centering
\includegraphics[width=0.5\textwidth]{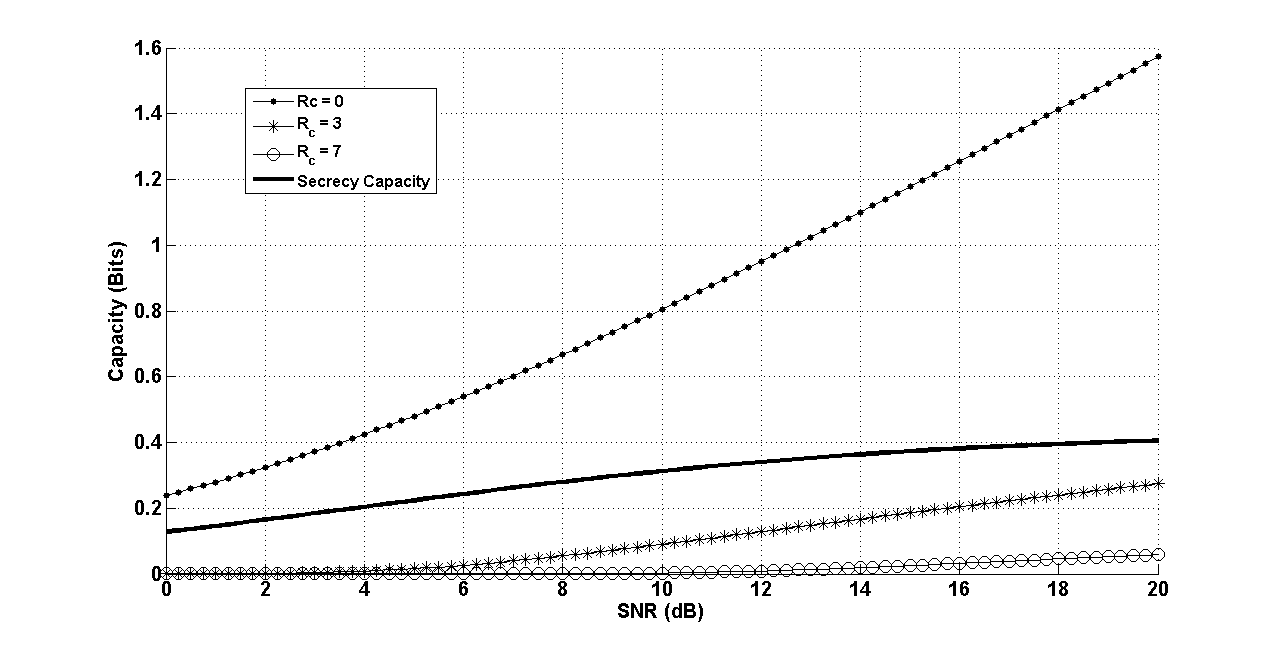}
\caption{$C_s$ and $C_e$ against SNR for $R_c =\left(0,3,7\right)$. \label{fig1}}
\end{figure}

Figure~\ref{fig1} gives the variation of $C_s$ and $C_e$ with SNR
under different constraints on the decoding capabilities of Eve.
It is clear from the figure that $C_e$ can be greater than $C_s$.
This can be the case for certain $R_c$ and SNR values. For
instance, in the case of $R_c=0$, if Eve receives the transmitted
packet with error, she discards it without any further attempts at
decoding. The instantaneous secrecy rate becomes $R_0$, which is
larger than that used in (\ref{sec_cap}) $C_s(i)= R_0 -
\log_2(1+h_e(i)P)$ where $C_s(i), h_e(i)$ are the instantaneous
secrecy rate, and Eve's channel power gain, respectively.
Averaging over all fading states, we can get a greater $C_e$ than
$C_s$. It is worth noting that, under the assumptions of the
symmetric scenario and the Rayleigh fading model, the scheme
proposed in~\cite{Poor} is not able to achieve any positive
secrecy rate.

\begin{figure}
\centering
\includegraphics[width=0.5\textwidth]{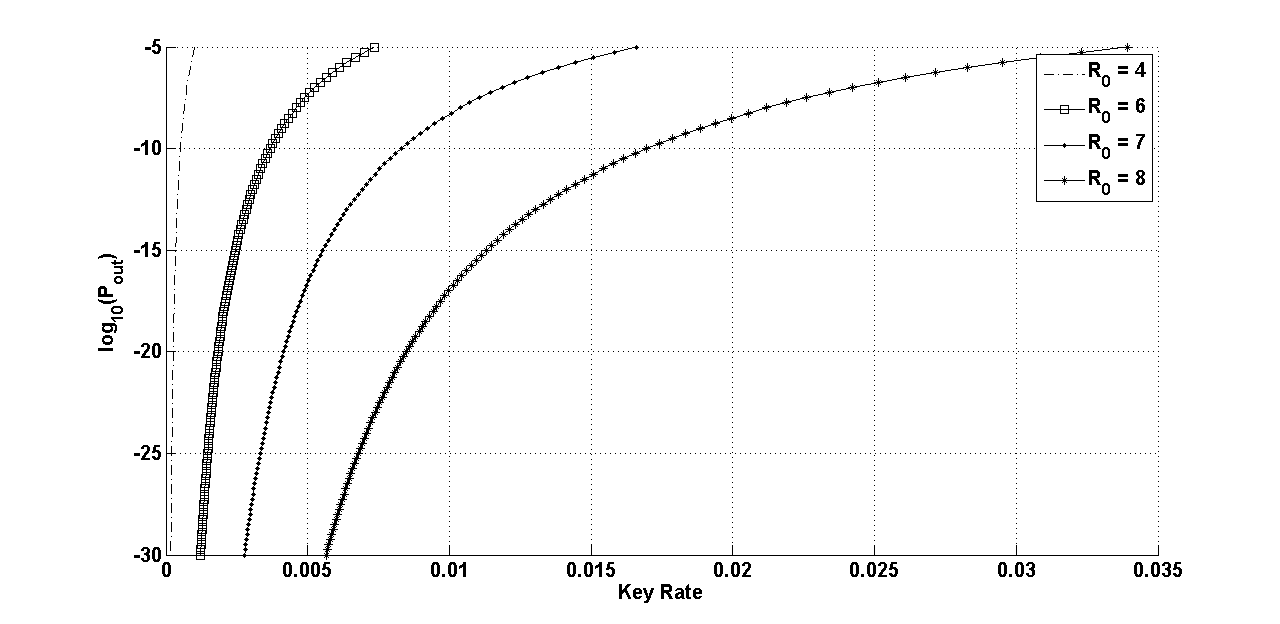}
\caption{Outage probability against key rate for $R_c=2$, $R_o=4$, $6$, $7$ and $8$, and an average SNR of $30$ dB. \label{fig2}}
\end{figure}

\begin{figure}
\centering
\includegraphics[width=0.5\textwidth]{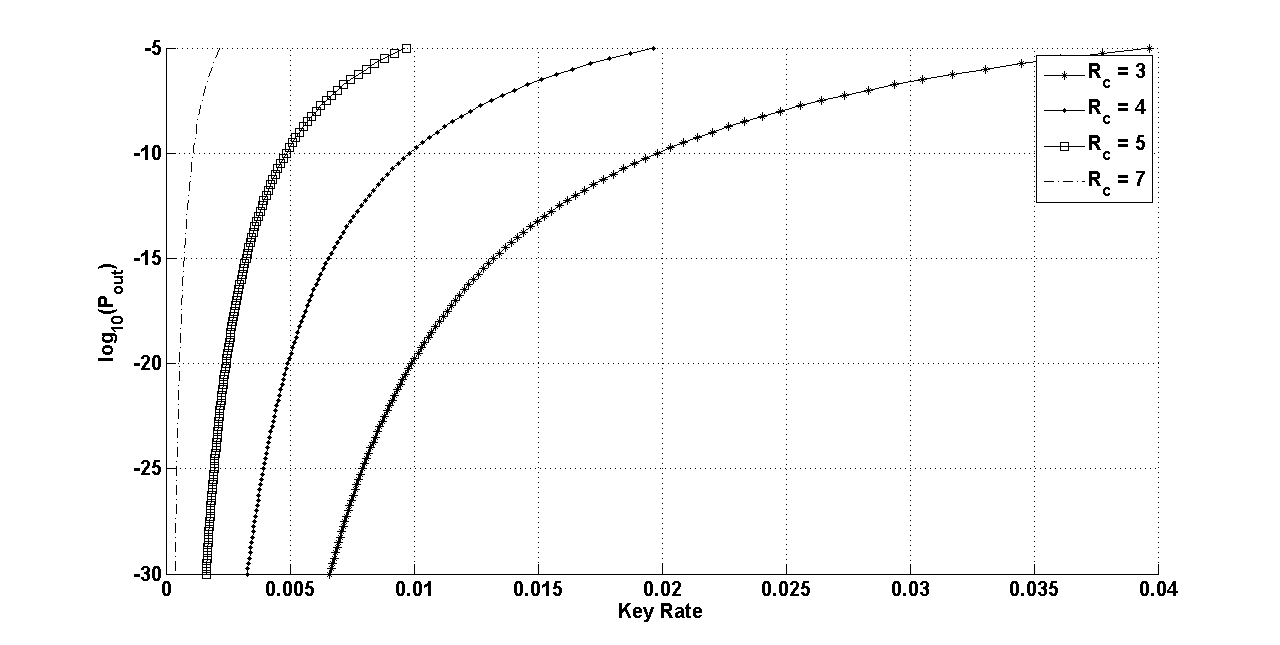}
\caption{Outage probability against key rate for $R_0=10$, $R_c=3$, $4$, $5$ and $7$, and  an average SNR of $30$ dB. \label{fig3}}
\end{figure}

\begin{figure}
\centering
\includegraphics[width=0.5\textwidth]{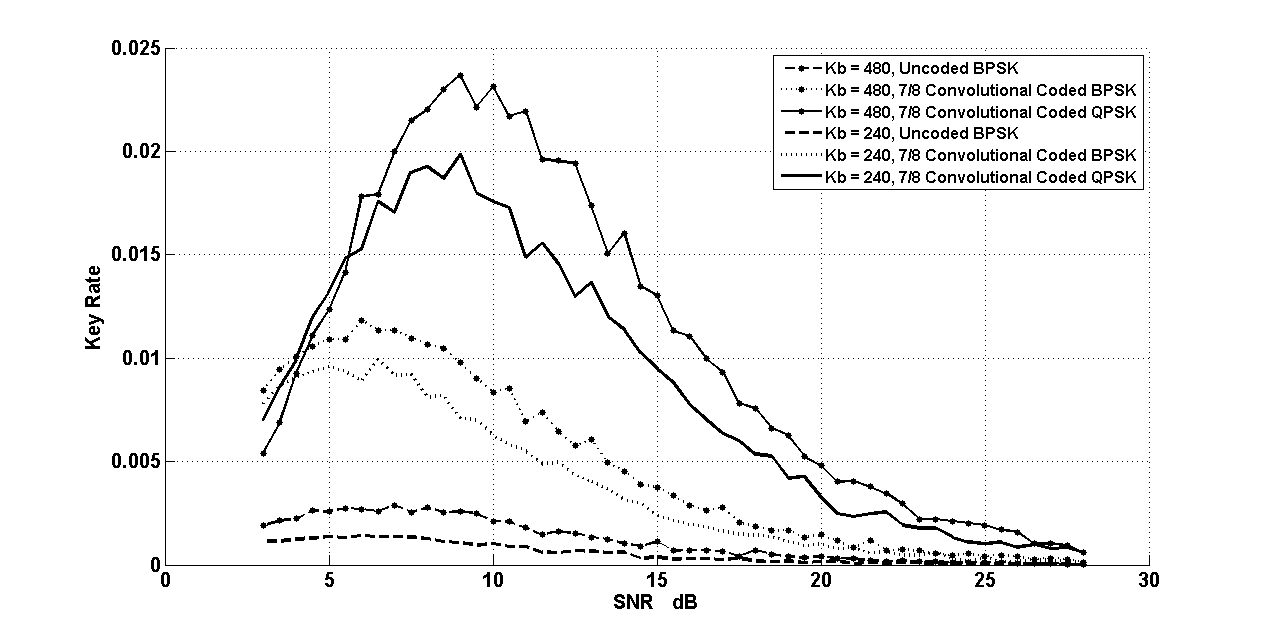}
\caption{The key rates required to obtain an outage of $10^{-10}$
against SNR for different packet sizes, $K_b=240$ and $480$ bits,
and different modulation schemes: uncoded BPSK, coded BPSK, and
coded QPSK. \label{fig4}}
\end{figure}

Next, we turn our attention to the delay-limited coding
constructions proposed in Section~\ref{sec:imp}.
Figures~\ref{fig2} and \ref{fig3} show, for different $R_0$ and
$R_{\rm c}$, the tradeoff between secrecy outage probability
versus key rate for the proposed rate $(k-1)/k$ coset secrecy
coding scheme assuming an optimal inner Gaussian channel coding.
Figure~\ref{fig2} gives key rate corresponding to a desired
secrecy outage probability, given some values for $R_0$ and $R_c$.
As is evident from Figure~\ref{fig3}, the key rate required to
obtain a certain outage probability gets smaller as $R_c$
increases. In Figure~\ref{fig4}, we relax the optimal channel
coding assumption and plot key rate for practical coding schemes
or no coding, and finite frame lengthes (i.e., finite $n_1$). The
code used in the simulation is a punctured convolutional code
derived from a basic $1/2$ code with a constraint length of $7$
and generator polynomials $133$ and $171$ (in octal). We assume
that Eve is Genie-aided and can correct an additional $50$
erroneous symbols (beyond the error correction capability of the
channel code). From the figure, we see that the key rate increases
with increasing SNR and then drops after reaching a peak value.
Note that we fix the transmission rate and make it independent of
SNR. A low SNR means more transmissions to Bob and a consequent
low key rate. As SNR increases, while keeping the transmission
rate fixed, the key rate increases. However, increasing SNR at
Eve's receiver means an increased ability to correctly decode the
codeword-carrying packets. This explains why the key rate curves
peak and then decay with SNR. Note also that for a certain
modulation and channel coding scheme, decreasing the packet size
in bits lowers the key rate. Reducing the packet size increases
the probability of correct decoding by Bob and, thus, decreases
the number of transmissions. However, it also increases the
probability of correct decoding by Eve and the overall effect is a
decreased key rate.

\section{Conclusions}\label{sec:con}
This paper developed a novel {\bf overlay} approach for sharing
secret keys using existing ARQ protocols. The underlying idea is
to distribute the key bits over multiple ARQ frames and then use
the authenticated ACK/NACK feedback to create a degraded channel
at the eavesdropper. Our results establish the achievability of
non-zero secrecy rates even when the eavesdropper is experiencing
a higher average SNR than the legitimate receiver. It is worth
noting that our approach does not assume any prior knowledge about
the instantaneous CSI; only prior knowledge of the highest average
SNR seen by the eavesdropper is needed. Moreover, we constructed a
low complexity secrecy coding scheme by transforming our channel
to an erasure wiretap channel which lends itself to explicit coset
coding approaches. Our theoretical claims were validated via
numerical examples that demonstrate the efficiency of the proposed
schemes. The most interesting part of our work is, perhaps, the
fact that it demonstrates the possibility of sharing secret keys
in wireless networks via rather simple modifications of the
existing infrastructure which, in our case, corresponds to the ARQ
mechanism.

\end{document}